\documentclass[preprint,3p,times,twocolumn,nonatbib]{elsarticle}

\usepackage[a-2u]{pdfx}
\usepackage[utf8]{inputenc}
\usepackage{lmodern}
\usepackage[T1]{fontenc}
\usepackage{xcolor}
\usepackage{stfloats}
\usepackage{paralist}
\usepackage{tabularx}
\usepackage{multirow}

\usepackage{microtype}
\usepackage{listings}
\definecolor{commentcolor}{rgb}{0.5,0.5,0.5}
\definecolor{darkgreen}{rgb}{0.09, 0.45, 0.27}
\definecolor{bgcolor}{rgb}{0.99,0.99,0.99}
\lstdefinelanguage{DeploymentDescriptor}{
    ndkeywords={kind,metadata,labels,name,spec,containers,app,template,probes,time,probability,ports,image,containerPort,probe,limits,timingRequirements},
    comment=[l]{\#}
}

\usepackage{flushend}

\begin{document}
    
\begin{frontmatter}

\title{Managing Latency in Edge-Cloud Environment\tnoteref{ttl1}}

\tnotetext[ttl1]{This is the author version of the article accepted for publication in Journal of Systems and Software.\\
© 2020. This manuscript version is made available under the CC-BY-NC-ND 4.0 license \url{http://creativecommons.org/licenses/by-nc-nd/4.0/}}

\author[add1]{Lubomir Bulej}
\ead{bulej@d3s.mff.cuni.cz}
\author[add1]{Tomas Bures\corref{cor1}}
\ead{bures@d3s.mff.cuni.cz}
\author[add1]{Adam Filandr}
\author[add1]{Petr Hnetynka\corref{cor1}}
\ead{hnetynka@d3s.mff.cuni.cz}
\author[add1]{Iveta Hnetynkova}
\ead{iveta.hnetynkova@mff.cuni.cz}
\author[add1]{Jan Pacovsky}
\author[add1]{Gabor Sandor}
\author[add2]{Ilias Gerostathopoulos}
\ead{i.g.gerostathopoulos@vu.nl}

\cortext[cor1]{Please address correspondence to Tomas Bures or Petr Hnetynka}
\address[add1]{Charles University, Faculty of Mathematics and Physics, Czech Republic}
\address[add2]{Vrije Universiteit Amsterdam, Department of Computer Science, Netherlands}

\begin{abstract}
Modern Cyber-physical Systems (CPS) include applications like smart traffic, smart agriculture, smart power grid, etc.
Commonly, these systems are distributed and composed of end-user applications and microservices that typically run in the cloud.
The connection with the physical world, which is inherent to CPS, brings the need to operate and respond in real-time. As the cloud becomes part of the computation loop, the real-time requirements have to be also reflected by the cloud.
In this paper, we present an approach that provides soft real-time guarantees on the response time of services running in cloud and edge-cloud (i.e., cloud geographically close to the end-user), where these services are developed in high-level programming languages.
In particular, we elaborate a method that allows us to predict the upper bound of the response time of a service when sharing the same computer with other services.
Importantly, as our approach focuses on minimizing the impact on the developer of such services, it does not require any special programming model nor limits usage of common libraries, etc.
\end{abstract}

\end{frontmatter}

\section{Introduction}
\label{sec:introduction}

Modern software systems and services are commonly distributed, composed of front-end applications running on end-user devices, and microservices running in the cloud.
This also holds for modern Cyber-physical Systems (CPS), such as data-driven applications for smart traffic, agriculture, or utilities.
These applications rely on data from sensors and perform computationally-intensive tasks (data analytics, optimization and decision making, learning and predictions) which cannot be executed on energy constrained devices and are therefore executed in the cloud.

However, the connection with the physical world inherent to CPS requires these systems to operate and respond in real-time, whereas cloud was primarily built to provide average throughput through massive scaling.
Real-time requirements impose bounds on response time, and when executing tasks in the cloud, a significant part of the end-to-end response time is due to communication latency.

The concept of \textit{edge-cloud} aims to reduce this latency by moving computation to a large number of smaller clusters that are physically closer to end-user devices.
Throughout the paper, we use the term edge-cloud in line with the definition of Satyanarayanan~\cite{satyanarayanan_emergence_2017}, i.e., we assume that computation which would be traditionally centralized in a data-center (in the case of a regular cloud), is moved to network edges, closer to the users.
This differs from the fog-computing (a related field of research), where the workload is traditionally decentralized, executing on end-user devices and a localized cloud (e.g., on an IoT gateway) is used for off-loading.

While usage of edge-cloud computing reduces communication latencies, edge-cloud alone does not guarantee bounded end-to-end response time, which becomes more determined by the computation time.
The reason is that the cloud itself focuses on optimizing the average performance and cost of computation, but does not provide any guarantees on the upper bound of the computation time of individual requests.
What is needed to address the requirements of modern cloud-connected CPS is an approach that can reflect the real-time requirements of modern CPSs even with the cloud in the computation loop.

Guarantees on a single request are the domain of real-time programming. But that itself is rarely a reasonable choice as it comes at a very high price of forcing developers to a low-level programming language, limited choice of libraries and the use of a relatively exotic programming model of periodic non-blocking real-time tasks.

In this paper, we advocate the use of standard cloud technologies (i.e., microservices packaged in containers running on top of Kubernetes) and modern high-level programming languages (e.g., Java, Scala, Python) for development of microservices that have real-time guarantees. 
We restrict ourselves to the class of applications where soft real-time requirements are enough (i.e., the guarantee on the end-to-end response is probabilistic –- e.g., \textit{in 99\% of cases the response comes in 100ms and in 95\% of cases the response comes in 40ms}). As it turns out this is a wide class of applications including augmented reality, real-time planning and coordination, video and audio processing, etc. Generally speaking, this class comprises any application that has a safe state and has a local control loop that keeps the application in the safe state while computation is done in the cloud. Consequently, the soft real-time guarantee pertains to qualities such as availability and optimality, but not to safety.

Also importantly, microservices of the considered class work with continuous workload (i.e., processing video or audio streams, etc.).
Starting and closing the workload (stream, etc.) are explicit operations.

In this context, the article presents an approach to providing soft real-time guarantees on response time of microservices running in a container-based cloud environment (e.g., Kubernetes), with microservices developed in high-level programming languages (Java in our case).

In particular, we elaborate a method that allows us to predict the upper bound of the response time (at a given confidence level) of a microservice when sharing the same computer with other microservices. This prediction method is essential for controlling admission to the edge-cloud and for scheduling deployment of containers to computers in the edge-cloud. Combined with adaptive control of deployment and re-deployment of components, this enables providing microservices with probabilistic guarantees on end-to-end response time.

An important feature of our approach is that we aim to remove the burden of specifying the required computational resources from the developer of services that need soft real-time guarantees. To this end, we treat microservices as black boxes and do not require any apriori knowledge about the microservices from the developer. Instead, our system performs experiments on the microservices to collect the data needed for performance prediction and deployment decisions.

In our approach, we are specifically targeting privately-controlled (non-public) edge-cloud environment, in which the edge-cloud operator controls not only the infrastructure but also the deployed microservices.
This contrasts with public clouds, in which the provider needs to cope with unknown applications, unknown workloads, unknown clients, etc.

The paper is structured as follows.
Section~\ref{sec:motivation} shows a motivation example. 
In Section~\ref{sec:core} we present our approach and in Section~\ref{sec:evaluation} its evaluation.
In Section~\ref{sec:discussion} we discuss limitations of our approach while Section~\ref{sec:related-work} shows related work.  Section~\ref{sec:conclusion} concludes the paper.

\section{Motivation example}
\label{sec:motivation}

As a motivation example application, we use a simplified yet realistic version of an augmented reality use-case taken from our ECSEL JU project FitOptiVis\footnote{\url{https://www.ecsel.eu/projects/fitoptivis}}, which focuses on developing an architecture for image- and video-processing for CPS.

\begin{figure}[h]
\includegraphics[width=\columnwidth]{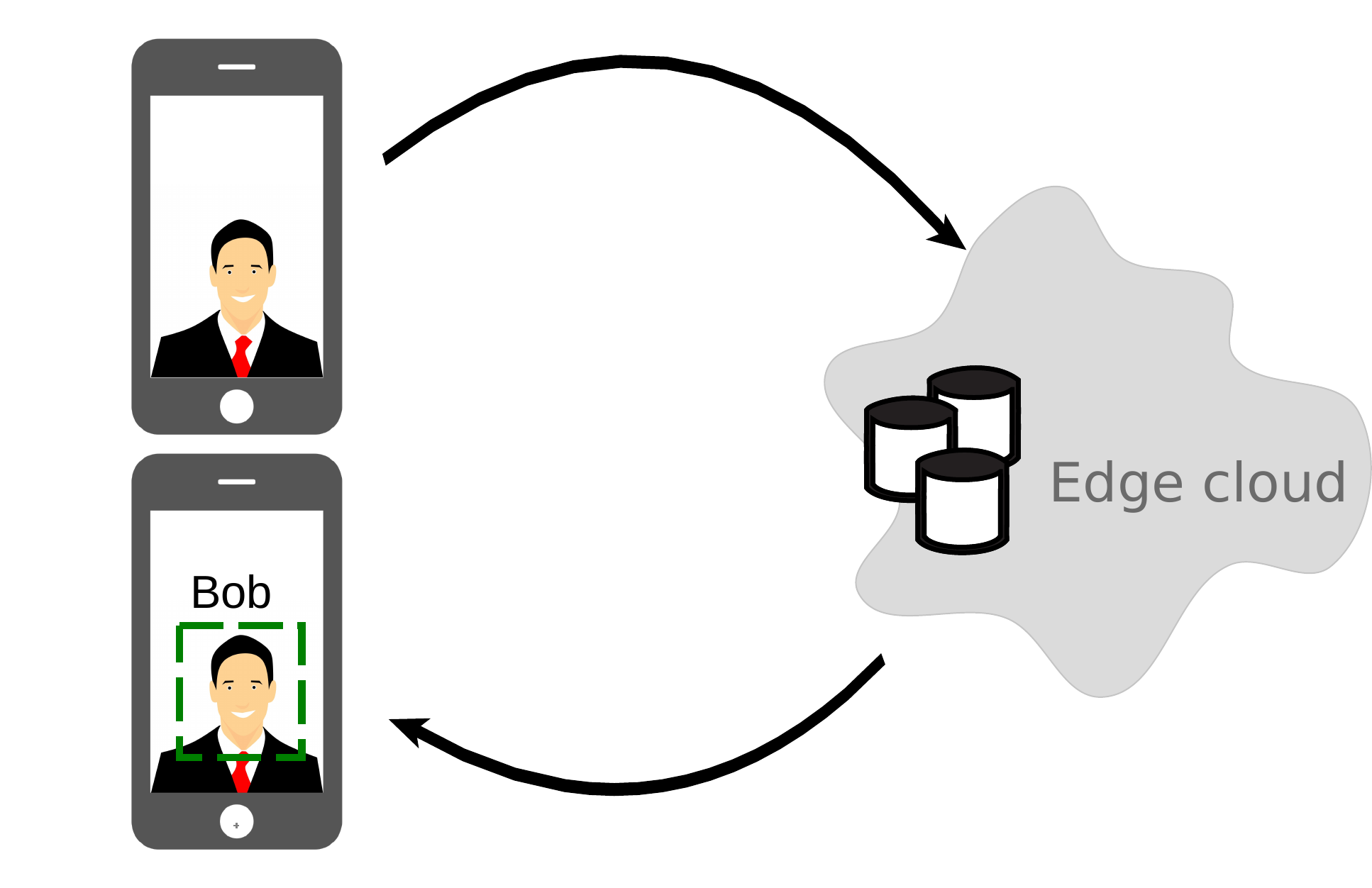}
\caption{Running example}
\label{fig:running-example}
\end{figure}

The example application (Figure~\ref{fig:running-example}) consists of a client application running on a mobile device (e.g., mobile phone) and a service hosted on 
edge-cloud nodes (close to the clients). 
The client application captures a video-stream (via the phone camera). The stream is sent to the service in edge-cloud for analysis and ``augmentation''. In our case, this comprises an identification of faces and lookup of names in a database. 

As a particular ``cloud'' technology, we are using Kubernetes (K8S)\footnote{\url{https://kubernetes.io/}}. Listing~\ref{lst:deployment-descriptor-orig} shows a piece of deployment descriptor (in the YAML language~\footnote{\url{http://yaml.org/}}) for the example application. The deployment descriptor captures the application's architecture as a set of microservices defined via Kubernetes's DeploymentSpec construct.

To work seamlessly, it is mandatory that ``augmentation'' information is received by the client without any significant delay and thus, in an ideal case, the service that does the analysis should be placed as close as possible to the device, i.e., in the edge-cloud, 
and the service in the edge-cloud should be collocated with other services (potentially from different tenants)
 such that it can still deliver the response in required time.
 
In this paper we focus on how to specify the latency requirements and the general approach to satisfy them. In particular, we address problem of which services in the edge-cloud can be collocated together such that edge-cloud resources are shared by several services and the latency requirements of each service in the edge-cloud are still met.

\begin{lstlisting}[float, caption=Deployment descriptor, label=lst:deployment-descriptor-orig, language=DeploymentDescriptor, escapechar=|, breaklines=true]
kind: Deployment
metadata:
  name: recognizer-deployment
  labels:
    app: recognizer
spec: # microservice specification
  template:
  metadata:
    labels:
      app: recognizer
  spec:
    containers:
    - name: recog
      image: repo/recog
      ports:
      - containerPort: 7777
\end{lstlisting}

\section{Managing Latency}
\label{sec:core}

In this section, we first outline our approach from the perspective of a developer of an edge-cloud application and present an overall architecture of the approach.
Then we describe an algorithm for predicting the response time upper bound of a microservice when colocated with other microservices.
Finally, we discuss the operational boundaries of the prediction algorithm.

\subsection{General strategy}
\label{sec:general_strategy}

The emphasis in our work is to minimize impact on the cloud-application developer. Essentially, in our approach, the cloud-application developer creates the typical K8S artifacts (code, container and deployment descriptor). The only extension brought by our approach is that the developer specifies real-time requirements (per service) in the K8S deployment descriptor.

These real-time requirements are interpreted by the cloud platform, which co-locates the microservices on the edge-cloud nodes in such a way that even though the microservices interact performance-wise, this performance interaction is below a threshold which would cause violation of the soft real-time requirements.

Contrary to the traditional existing means of cloud deployment, the developer does not have to deal with selection of VM type, number of virtual CPUs, memory, IOPS, etc. Similarly, the developer does not have to specify any auto-scaling rules (including triggers). We believe that these concerns are internal to the cloud and that the developer should not be confronted with these questions because anyway he/she does not typically have a well-justified answer supported by experiments.

As such, we work with the clean abstraction where the developer provides the application and the soft real-time requirements. The responsibility to asses performance of the cloud application (including required number of virtual CPUs, memory, IOPS, etc.) lies on our edge-cloud platform. In our approach, the deployment in the edge-cloud happens as follows:
\begin{inparaenum}[(1)]
\item The developer develops the application as a collection of microservices and creates the deployment descriptor where he/she specifies the real-time requirements.
\item The developer submits the application to the edge-cloud. 
\item The edge-cloud performs performance assessment that consists of performance tests of the microservices to determine their baseline performance and their performance when co-located with other workloads. 
This assessment answers the question whether the soft real-time requirements can be guaranteed at all and what deployment parameters (in terms of virtual CPUs -- shared or dedicated, memory, reserved IOPS) must be allocated for the microservice. 
\item After the assessment is done, the edge-cloud informs the developer (potentially including the price for the service). 
\item The developer confirms the deployment. 
\item The edge-cloud calculates and performs the deployment using the measured data from the assessment.
\end{inparaenum}

To capture the soft real-time requirements on latency, we extend the K8S deployment descriptor by additional section attached to specification of a microservice. Such an extended deployment descriptor is shown in Listing~\ref{lst:deployment-descriptor}. 

Technically, the soft real-time requirements are captured by the \textit{probes} part of the deployment descriptor (starting at line~\ref{inlst:probes}).
A probe is a special function (provided by the application developer) allowing for performance measurements. The real-time requirements are specified as statistical expressions over the probes as shown in the example. 

We assume that these probes are developed in a way that a probe ``emulates'' a typical behavior of one request for a particular actual microservice. The probe has no inputs -- it is a self-contained performance test method that encapsulates the workload (e.g., an image for the image recognition service) that is needed to execute the performance test. Another important point is that a probe can be executed without affecting the functionality (in particular the state) of the microservice -- this allows us to execute the probes even in a production environment to asses if a microservice still complies to the real-time requirements.

By connecting the soft real-time requirements to the probes, we create a kind of SLA (Service-level Agreement) between the developer and the cloud platform that can be autonomously measured by the cloud platform. 

This way, the probes and the requirements over them provide a simple and feasible contract between the developer of a latency sensitive microservice and our method for scheduling workload in the edge-cloud. Thanks to the probes, our approach can automatically assess the service while treating it as a black box. 
This can be understood in fact as an automated creation of SLA rules and is one of the key novelties of our approach. 

\begin{lstlisting}[float, caption=Deployment descriptor, label=lst:deployment-descriptor, language=DeploymentDescriptor, escapechar=|, breaklines=true]
kind: Deployment
metadata:
  name: recognizer-deployment
  labels:
    app: recognizer
spec: # microservice specification
  template:
  metadata:
    labels:
      app: recognizer
  spec:
    containers:
    - name: recog
      image: repo/recog
      ports:
      - containerPort: 7777
      probes:   # probes |\label{inlst:probes}|
      - name: recognize
        timingRequirements:
      - name: recognize limit
        probe: RecognizePeople
        limits:
        - probability: 0.99
          time: 100 # Max. 100ms in 99% cases
        - probability: 0.95
          time: 40 # Max. 40ms in 95% cases
\end{lstlisting}

Taking a look at the motivation example, the application consists of a client part (capturing a video stream) and the microservice detecting and recognizing the faces plus a database microservice with for name-face lookup.
The database microservice is not important now, as there are no timing requirements placed on it (the recognizer loads all necessary data from it at its start). As such, we leave out the database for further discussion.

To allow specification of soft real-time requirements over the \textit{recognizer} service, the developer provides a \textit{probe} that utilizes the same procedure (as the regular API of the microservice) for detecting and recognizing a predefined face in a predefined stream. Nevertheless the probe does not affect the state of the microservice (e.g., the cumulative count of detected faces).

To assess whether a microservice can be deployed, all probes of the microservice are exercised both in isolation and with predefined background workloads to characterize the microservice's resource demand along three dimensions---CPU, memory, and IOPS bandwidth.
This allows the edge-cloud infrastructure to build a model of the microservice which is then used to decide whether the service can be admitted and where to deploy it so that the timing requirements can be met.

We exclude network from this characterization because we assume that the latency sensitive traffic on the network will be by an order of magnitude below the overall capacity of the network and that the latency sensitive network can be configured with a dedicated QoS class that will give it priority over regular traffic on the network. As such, we can consider the network (in connection to the latency sensitive microservices) as a virtually unlimited resource.

\subsection{Overall architecture}
\label{sec:architecture}

\begin{figure*}[!t]
	\begin{minipage}{\textwidth}
		\centering
		\includegraphics[width=\textwidth]{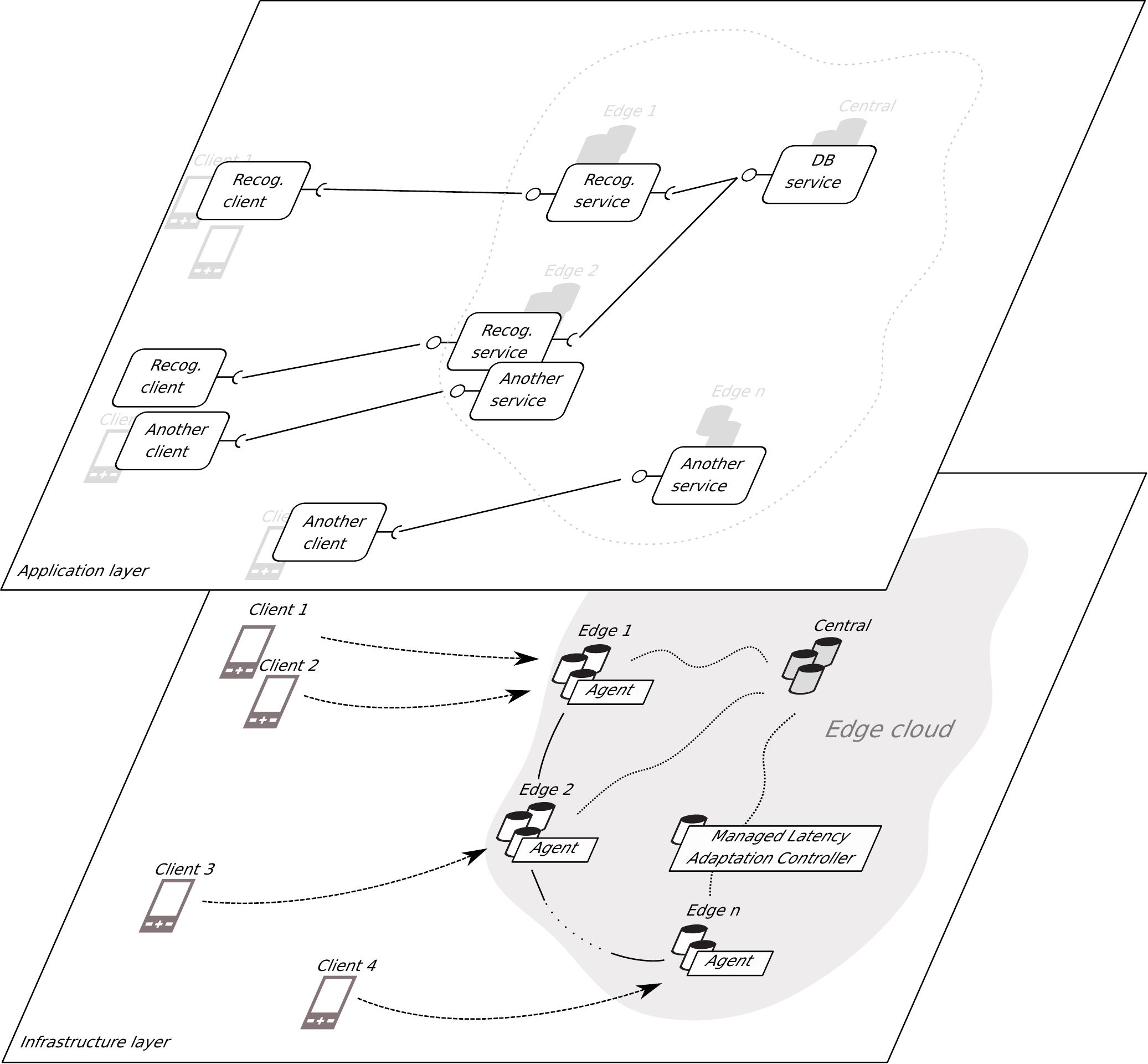}
		\caption{Overall architecture}
		\label{fig:architecture}
	\end{minipage}
\end{figure*}

The overall edge-cloud architecture considered of our approach (while running the motivation example) is depicted in Figure~\ref{fig:architecture}.
There are two views on the architecture depicted~--~the infrastructure layer (i.e., how the computers of the edge-cloud are organized) and the application layer (i.e., microservices running on the computers).

On the infrastructure layer, the edge-cloud architecture consists of several Edge data centers (Edge 1, Edge 2, etc.~--~each of them consisting of several computers), where microservices with timing requirements are deployed.
Additionally, there is Central data center, where microservices without requirements are deployed.
The \textit{managed latency adaptation controller} (in addition to be a management interface of the edge-cloud) manages the deployment and re-deployment at individual data centers through \textit{agents} located at each center.
Clients are connected to microservices deployed to edge data centers, typically to the closest one (but it depends on actual utilization in the centers).
Each client can run several different applications depending on different microservices.

Because we are targeting a privately controlled edge-cloud, we assume the edge data centers to be largely homogeneous, i.e., composed of computers with similar or identical configuration.

On the application layer, the figure shows a possible deployment of two applications.
The Client 1 and 3 run the motivation example, which has the recognizer microservice deployed in the edge center and the database microservice (without timing requirements) running in the central center.
Client 3 and 4 runs another application (which, in contrast to the motivation example, has only a single microservice).

To ensure that the timing requirements are met at run-time, we use the adaptive systems architectural pattern where the \textit{managed latency adaptation controller} manages a K8S edge-cloud (consisting of multiple data centers). The managed latency adaptation controller interprets the soft real-time requirements specified as part of the extended deployment descriptor. It decides on how to collocate the microservices and instructs an existing (unmodified) K8S edge-cloud to perform such deployment. The interaction between the managed latency adaptation controller and the K8S edge-cloud is realized using the standard K8S API.

The managed latency adaptation controller implements a MAPE-K self-adaptation loop~\cite{kephart_vision_2003}, which periodically checks whether the soft real-time requirements are met and using \textit{Knowledge} predicts near-future requirements. 
This prediction enables not only intervention after the detection of requirements violation but also, more importantly, proactive reaction ahead of time if needed. 

The same adaptation loop is used to manage initial deployment but also re-deployment of microservices.
Re-deployment is in fact nearly identical to initial deployment~--~calculation of real-time requirements is done periodically within the loop and takes into account the current placement of microservices to prevent unnecessary re-locations.

In particular, in the \textit{Monitoring} phase of the loop, the controller periodically monitors the probes of the currently deployed microservices and the resource utilization on the computers in the edge-cloud cluster.
Then in the \textit{Analysis} phase, the controller uses the collected data and the microservice models to perform a what-if analysis (using a prediction algorithm described shortly) to evaluate deployment alternatives for the microservices that are currently running and the microservices submitted for deployment.
If a (better) fitting deployment is found, then the \textit{Planning} phase comprises preparing low-level deployment tasks for the selected alternative, which are then carried out during the \textit{Execution} phase to re-deploy the microservices to satisfy the timing requirements.

On each node, the information about a microservice obtained during the assessment phase is used to assign each deployed microservice the resources needed to perform its tasks within the timing constraints.
This resource allocation is strictly enforced using features of the operating system, containerization technology, or the virtualization platform.
This is necessary to prevent microservices from exceeding their allocated share of resources (due to, e.g., a sudden spike in the number of clients), which could have a negative impact on the execution time of other microservices.
In our prototype, we enforce the resource allocation using features of Docker and Linux cgroups.

\subsection{Performance prediction of collocated workloads}

For the analysis of deployment alternatives, we introduce a novel performance prediction algorithm.
It is based on a statistical characterization of measurements followed by similarity comparison, revealing dependencies between background workloads (i.e., microservices).
From measurements, we first build a structured complex data set. Each time a performance prediction of a particular scenario is needed,  relevant prediction data are extracted into a linearized data fitting model. This model is then solved by a constrained least-squares method, giving a reliable order statistics estimate of the performance, including its fidelity.

First, we explain how initial data are collected. Assume we have $n$ initial workloads $A_i, i=1, \dots, n$ and $l$ parameters (random variables) characterizing their behavior that we wish to measure (such as run time).
For each $A_i$, $p_i$ scenarios of its run with other background workloads are selected and measured. To ensure robustness of the predictor, a fixed number of $k$ measurement repeats under various conditions are realized, representing finally a random sample of the length $k$ for each of the $l$ random variables. The scenarios include a single test measurement for each of the workloads $A_i$ (i.e., a scenario without any other background workload) and various selected combined test. Denote $m = \sum_{i=1}^n p_i$ the total number of tested scenarios. To simplify the exposition, we use below the standard MATLAB tensor notation (rather than multidimensional algebra notation) for selecting its particular subblocks.
The initial data set is then represented by three data structures:
\begin{itemize}
    \item $S \in {\cal R}^{k \times l \times m}$: A three-way tensor, where each of the frontal slices (i.e., a $k \times l$ matrix $S(:,:,r)$) corresponds to the same scenario with the rows being the repeated measurements. The frontal slices are organized such that the $p_i$ scenarios
    corresponding to the same workload $A_i$ represent a subtensor $S(:,:,q:q+p_i)$ with the single test being its first frontal slice.
    \item $M \in {\cal Z_{+}}^{n \times m}$: A matrix with the entries being positive integers (including zero). The $i$-th column  of $M$ encodes the scenario corresponding to the $i$-th frontal slice of $S$. Here $M(i,j)=0$ in case of the $i$-th processes being not included
    in the $j$-th scenario, while $M(i,j) \neq 0$ gives the number of background runs otherwise.
    \item $v \in {\cal Z_{+}}^{n+1}$: A vector with the entries being positive integers. It represents a compressed row encoding of organization of scenarios in the tensor $S$ allowing for fast access to relevant frontal slices during the computation. Here $v(1)=1$, while
    $v(i+1)-v(i) = p_i$ for $i=2,\dots,n$. Then, e.g., $S(:,:,v(i)), i=1,\dots,n$ are the single test frontal slices of $S$.
\end{itemize}

\begin{figure*}[!t]
    \centering
    \begin{minipage}{0.8\textwidth}
        \includegraphics[width=\textwidth]{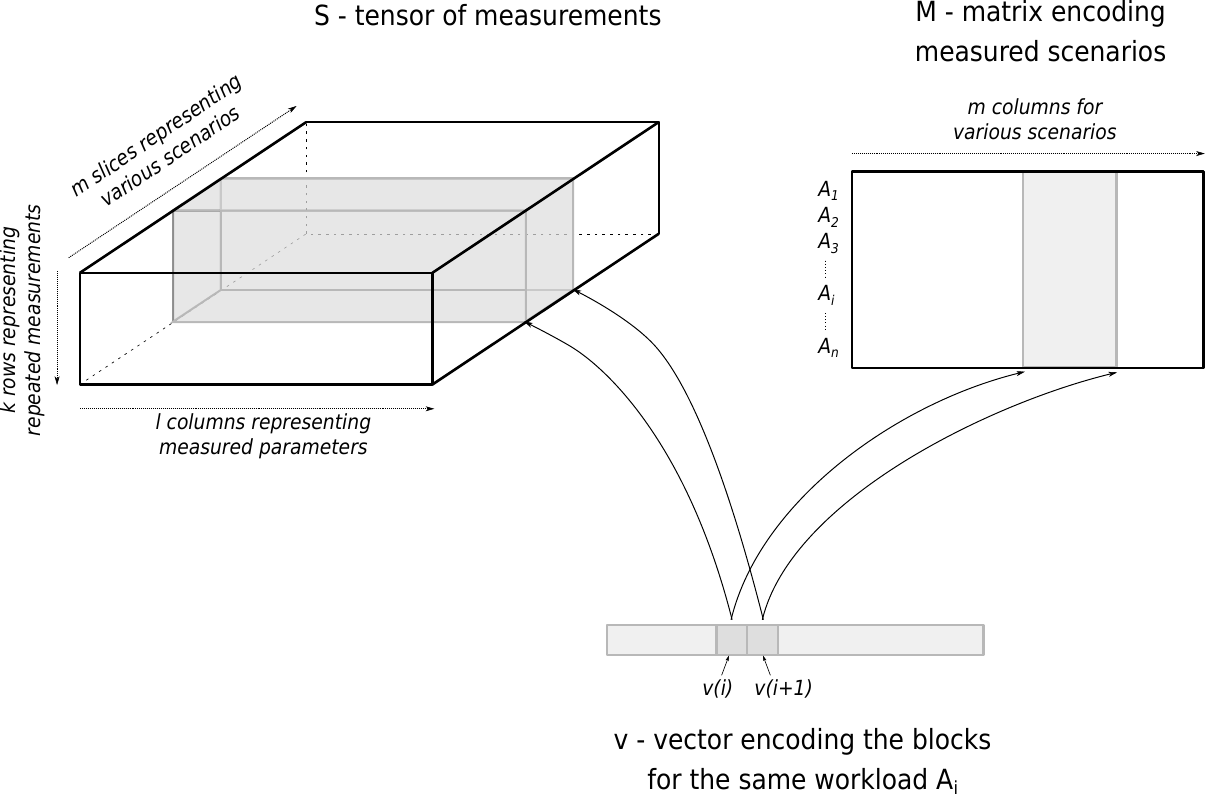}
        \caption{Initial data collection for the predictor}
        \label{fig:alg-01}
    \end{minipage}
\end{figure*}

\begin{figure}[h]
    \includegraphics[width=\columnwidth]{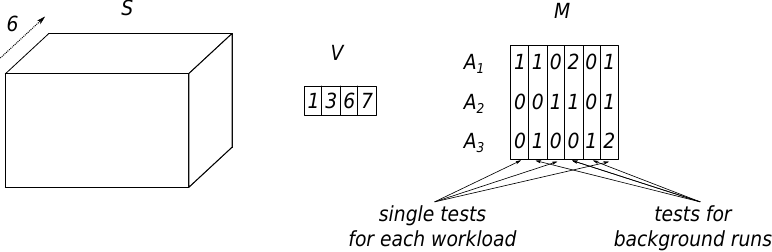}
    \caption{Example of data set for three workloads $A_1$, $A_2$, $A_3$}
    \label{fig:alg-02}
\end{figure}

Figure~\ref{fig:alg-01} illustrates the data set. The gray blocks in $S$ and $M$ represent data corresponding to measuring performance of the same workload $A_i$. The position of the blocks is encoded by the entries of the vector $v$, the block in $S$ is formed by frontal slices
on the positions from $v(i)$ to $v(i+1)-1$, similarly for the columns of $M$. Figure~\ref{fig:alg-02} gives an example of initial data for three
workloads, where for each of them  one single test (odd columns of $M$, odd frontal slices of $S$) and one combined test (even columns and slices) is available.

Having the initial data $S,M$ and $v$, the prediction approach consists of three phases:
\begin{itemize}
    \item data preprocessing (evaluation of statistical data representation),
    \item task fitting (constraint linear regression for a positive integer problem),
    \item data-based prediction (weighted combination of predicted dependencies).
\end{itemize}
The first phase represents all computations that can be done apriori to save the computational costs in later phases.\\ 

\textbf{PHASE 1:} Here the goal is to evaluate several statistical characteristics (for each of the given scenarios) in order to capture
dependencies of the $l$ parameters of interest on the measurement conditions. We proceed as follows. We construct a three-way tensor
$Sstat \in {\cal R}^{\tilde{k} \times l \times m}$, where the frontal slice $Sstat(:,:,r)$, $1 \leq r \leq m$ always corresponds to 
the $r$-th scenario.
The first rows in $Sstat(:,:,r)$ will contain information about statistical distribution of the measurements.
In particular, we compute the sample mean and median value, selected sample percentiles (e.g., $90\%$), standard and relative deviation, standard error, and the difference between the sample maximum and sample minimum value, i.e. (for each column $j$, $j=1,\dots,l$):\\

\noindent $Sstat(1,j,r) = \frac{1}{k} \sum_{i=1}^k S(i,j,r)$  \\
$Sstat(2,j,r) = \mbox{MED}(S(:,j,r))$\\
$Sstat(3,j,r) = 90^{th}\mbox{PERC}(S(:,j,r))$\\
$Sstat(4,j,r) = \sqrt{\frac{1}{k-1} \sum_{i=1}^k (Sstat(1,j,r) - S(i,j,r))^2}$ \\
$Sstat(5,j,r) = Sstat(4,j,r)/Sstat(1,j,r)$ \\
$Sstat(6,j,r) = Sstat(4,j,r)/\sqrt{k}$\\
$Sstat(7,j,r) = \max_{i=1,...,k}S(i,j,r) - \min_{i=1,...,k}S(i,j,r)$ \\

While the characteristics such as mean or median indicate some typical behavior, the sample maximum and minimum capture information about the extreme measurements.
Their difference is used later for effective penalization of measurements with lower fidelity, ensuring reliability of the performance prediction. Alternatively,
one could determine here prior bounds for the confidence intervals of the sample percentiles of interest.

The last block of rows in $Sstat(:,:,r)$ contains various quantities allowing to reveal the dependencies between the performance of workloads $A_i, i=1, \dots, n$.
These include in particular slowdown parameters corresponding to the difference between the sample percentile of the random sample in case of $A_i$ running separately (single test) and running with other background workloads (general scenario).
E.g., the absolute and relative version for $90^{th}\mbox{PERC}$ is given by \\

\noindent $Sstat(8,j,r) = Sstat(3,j,r) - Sstat(3,j,s)$ \\
$Sstat(9,j,r) = Sstat(8,j,r)/ Sstat(3,j,s)$, \\

\noindent where for a particular $r$ we find $i$ such that $v(i) \leq r < (v(i+1)-1)$ and put $s = v(i)$. Similarly, influence parameters can be determined representing influence of background workloads on the estimated lower and upper bounds of the confidence intervals. \\ 

\textbf{PHASE 2:} Having the initial data $S, M, v$, their statistical characteristics $Sstat$ and a user-specified prediction requirement (i.e., a question), we first need to detect precomputed scenarios relevant for the prediction. We allow two types of scenario questions:
\begin{itemize}
    \item Q1: performance prediction for one of the already tested workloads $A_i$, $i \in \{1, \dots, n\}$,
    \item Q2: performance prediction for a new workload $A_{n+1}$ with available single test.
\end{itemize}

For Q1, the situation is simpler.
The prediction must be based on the statistical characteristics of the scenarios for $A_i$. In order to build the prediction model, we extract the ${\tilde p}_i = p_i-1$ columns of the integer matrix $M$ (encoding the scenarios for the slices of $S$) corresponding only to the workload $A_i$, except the column for the single test, i.e.,
$$
M_1 = M(:, (v(i)+1):(v(i+1)-1)).
$$
In Figure~\ref{fig:alg-01} this corresponds to the gray block in $M$, except of its first column.
The questioned scenario is represented by an integer vector $b \in {\cal Z_{+}}^{n}$ with meaning analogous to that of a column of $M$, i.e., 
we put $b(i) =0$ in the case of the workload $A_i$ not being included in the scenario and $b(i) \neq 0$ otherwise. 
This gives us the data fitting problem
$$
M_1 x \approx b
$$
modeling the unknown correlation between the question $b$ and the preselected initial scenarios $M_1$. Such data fitting problem
can be solved by various methods. In order to reduce underestimating in PHASE 3, it is necessary to ensure that the entries of $x$ are not negative. For high-dimensional sparse models with a large number of measurements, it has been shown recently (see \cite{slawski2013})  that
constraint least-squares method with nonnegative constraint (NLS) is advantageous over other standard approaches. Modifications of other methods including nonnegative least-mean-square algorithm (see \cite{CHEN2016131}) are also applicable. We determine $x$ by solving the NLS problem
$$
\min_{x \in {\cal R}^{{\tilde p}_i}} \|b-M_1 \, x\| \quad \mbox{subject to} \quad 0 \leq x(i), \ \ i = 1, \dots, {\tilde p}_i,
$$
where $\|.\|$ is the Euclidean norm (other norms such as weighted ones can be also considered).
Then the nonzero entries of $x^{NLS}$ determine which frontal slices of $Sstat$ are relevant for the prediction and the value $x^{NLS}(i)$ itself represents a positive weight (i.e., importance) of the corresponding slice.
Note that if $x = 0$ then the model $M_1$ does not match the question $b$ meaning that not enough measurements are available for the prediction.
This situation has to be solved separately either by including an extra measurement into $S$ (if available) or by modifying Q1 to Q2, where the single test for $A_i$ is removed from further processing of this particular task to avoid cycling.

For Q2, the prediction is based on detecting a workload among the tested ones, i.e., $A_j$, $j=1, \dots, n$, that most resembles the new $A_{n+1}$.
Having in hand the new single test $Snew \in {\cal R}^{k \times l}$ for $A_{n+1}$, we compute its characteristics $Sstatnew\in {\cal R}^{\tilde{k} \times l}$ similarly as we did for the initial data. Then we compare selected characteristics from $Sstatnew$ with the frontal slices of $Sstat$ corresponding to single tests.
Considering some similarity measure $f$ between selected sample characteristics, we are looking for a workload $A_{i}$ such that
$$
A_{i} = \arg\min_{j=1, \dots, n}f(A_j,A_{n+1}).
$$
One can choose here for example a carefully selected weighted vector norms of the difference between mean, median and deviation for most relevant measured random variables (i.e., columns of $Sstat$). Now we proceed as in Q1 with $A_{n+1}$ replaced by the workload $A_{i}$. The tensors $S, Sstat$, the matrix $M$ and the vector $v$ are extended by the new measurement.  \\

\begin{figure}[h]
    \centering
    \includegraphics[width=0.8\columnwidth]{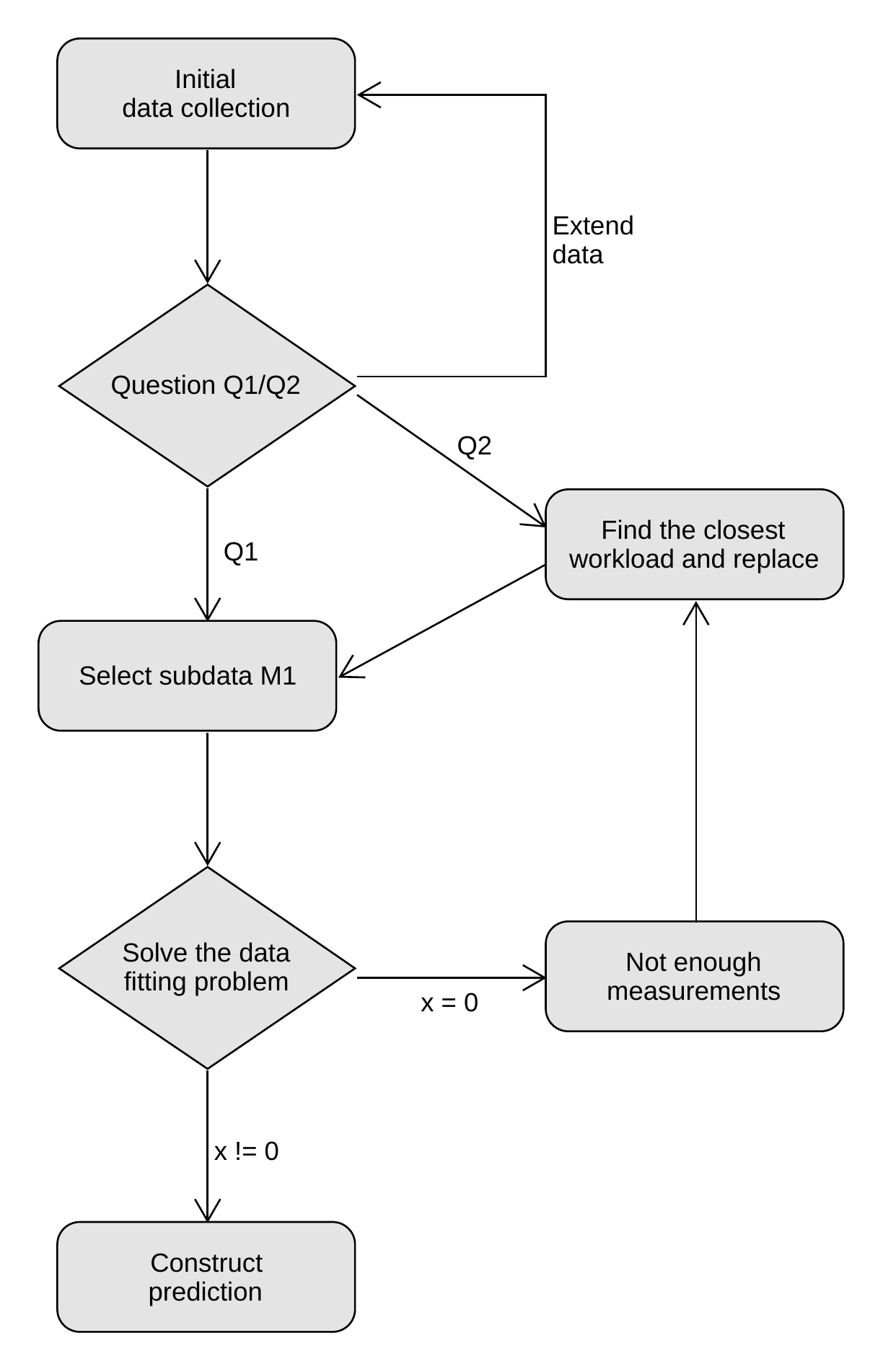}
    \caption{Algorithm overview}
    \label{fig:alg-seq}
\end{figure}

\textbf{PHASE 3:} Based on the approximation $x^{NLS}$, a weighted combination of workload dependencies (saved in $Sstat$)
can be used to predict the behavior of the scenario from the question Q1 or Q2.
Recall that the columns of the matrix $M_1$ encode the only relevant scenarios detected in PHASE 2 with their importance represented 
by the nonnegative entries of $x^{NLS}$.
Thus we first extract ${\tilde p}_i$ frontal slices of $Sstat$ corresponding to $M_1$. Then we weight them by $x^{NLS}$ as follows
$$
Rstat_j = \alpha_j * x^{NLS}(j) * Sstat(:,:,(v(i)+j)), 
$$
$j = 1, \dots, {\tilde p}_i,$ where $\alpha_j > 0$ is a safety penalization coefficient
(determined by scaled difference between the sample maximum and minimum measurement value).
Note that each resulting $Rstat_j \in {\cal R}^{\tilde{k} \times l}$ is a matrix.
A characteristic of interest for the scenario Q1 or Q2 can then be predicted from $Rstat_j, j = 1, \dots, {\tilde p}_i$, where the key role is played by the last block of rows representing slowdown parameters for the percentiles.
For example, an estimate of the $90^{th}\mbox{PERC}$ of the expected performance for Q1 is obtained by shifting the percentile for the single test of $A_i$ (saved in $Sstat(3,:,v(i))$) by a linear combination of estimated weighted slowdowns as follows
$$
90^{th}\mbox{PERC}(Q1) \approx  Sstat(3,:,v(i)) + \sum_{j=1}^{{\tilde p}_i} Rstat_j(8,:).
$$
The whole prediction algorithm is summarized in the schema in Figure~\ref{fig:alg-seq}.

\subsection{Ensuring operational boundaries}
\label{subsec:operational_boundaries}

The interactions among colocated microservices sharing the underlying physical resources are generally complex, and often non-linear---especially when the physical resources are nearing exhaustion.
Consequently, the prediction accuracy varies with different combinations of applications and resources used, and cannot provide actionable results for all possible scenarios.

To ensure that the predictor can be used with confidence within the adaptation loop, it is critical to establish the predictor's operational boundaries and ensure that the managed system stays within the boundaries.
The boundaries can be expressed as limits on the utilization of the CPU, memory, and IO resources used to characterize microservice workloads.

To demonstrate the predictor and its limits along the three dimensions, we use a synthetic benchmark tool to execute colocated workloads targeting a particular resource (CPU, memory, IO bandwidth) at varying levels of utilization and compare the predicted and the actual benchmark run times.
All experiments were carried out on a 64-bit quad-core Intel system\footnote{Intel Xeon E3-1230v6 @ 3.50GHz, \url{https://ark.intel.com/products/97474/Intel-Xeon-Processor-E3-1230-v6-8M-Cache-3-50-GHz-}} running Fedora Linux 28\footnote{Kernel 4.17.3-200.fc28.x86\_64; Docker 18.03.1-ce; OpenJDK 1.8.0\_191}.
Hyper-threading, turbo-boost and other power-management features were disabled to obtain stable timing results.

\subsubsection{CPU utilization}
In this experiment we focus on the interactions between colocated CPU-bound microservices, and analyze how their response times change as their demand for CPU increases and exceeds the available CPU capacity.
To this end, we evaluate a scenario in which we deploy 5 CPU-bound microservices to 5 containers.
Each container is limited to 25\% of the overall CPU capacity, i.e., one core of the experimental quad-core machine.
We simulate the microservices using our benchmark tool, and vary the target CPU utilization of each service in the range [0\%, 5\%, 10\%, .., 100\%] (at 0\% utilization the benchmark only passively waits, while at 100\% utilization it only performs CPU-intensive calculations), which results in overall CPU demand ranging 0\% to 125\%.

\begin{figure}
    \includegraphics[width=\linewidth]{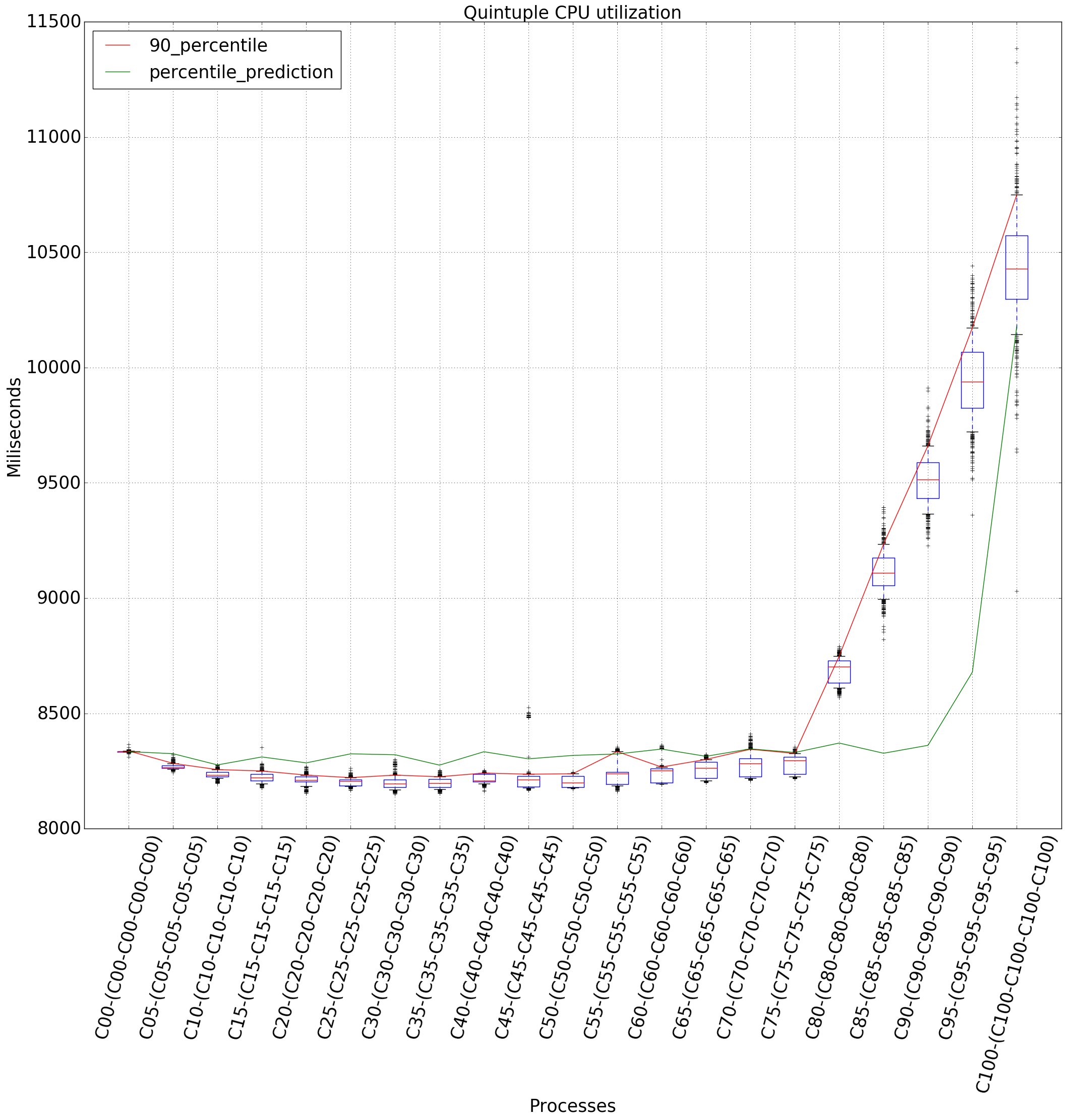}
    \caption{Benchmark response times for 5 colocated microservices with varying levels of CPU utilization.}
    \label{fig:cpu_paper}
\end{figure}

The results of the experiment are shown in Figure~\ref{fig:cpu_paper}.
Points on the X axis correspond to CPU utilization in each of the 5 containers, while the Y axis shows the response time.
The blue boxplots represent observed response times for each container CPU utilization setting, with the red line connecting 90th percentiles of the observed response times.
The response times remain stable until 75\% container CPU utilization (94\% overall), and start to increase when the container CPU utilization reaches 80\% (100\% overall).
The response time grows linearly with the CPU overcommitment percentage, and at 100\% container CPU utilization (125\% overall) the response time is roughly 125\% of the value observed while the overall CPU capacity was sufficient, which matches the intuition.

In contrast, the predicted response time 90th percentile (green line) starts increasing only after the container CPU utilization reaches 95\% (119\% overall), which is too late.
Because the model does not specifically account for context switching and other system overheads consume a variable portion of the overall CPU capacity, we should limit the predictor's operational boundaries to less than 100\% overall CPU utilization (94\% in this case).

\subsubsection{IO throughput}
In this experiment we focus on the interactions between IO-bound microservices accessing a shared disk, and analyze how their response times change as their demand of IO bandwidth increases and exceeds the available IO capacity.
We define the available IO capacity as the maximal amount of data which a computer can read/write per second, and IO utilization as a percentage of this capacity used by a microservice.
This time we deploy 3 microservices simulated by our benchmark tool to 3 containers, and vary their IO utilization from 0\% to 85\%, resulting in overall IO demand ranging from 0\% to 255\%.

\begin{figure}
    \includegraphics[width=\linewidth]{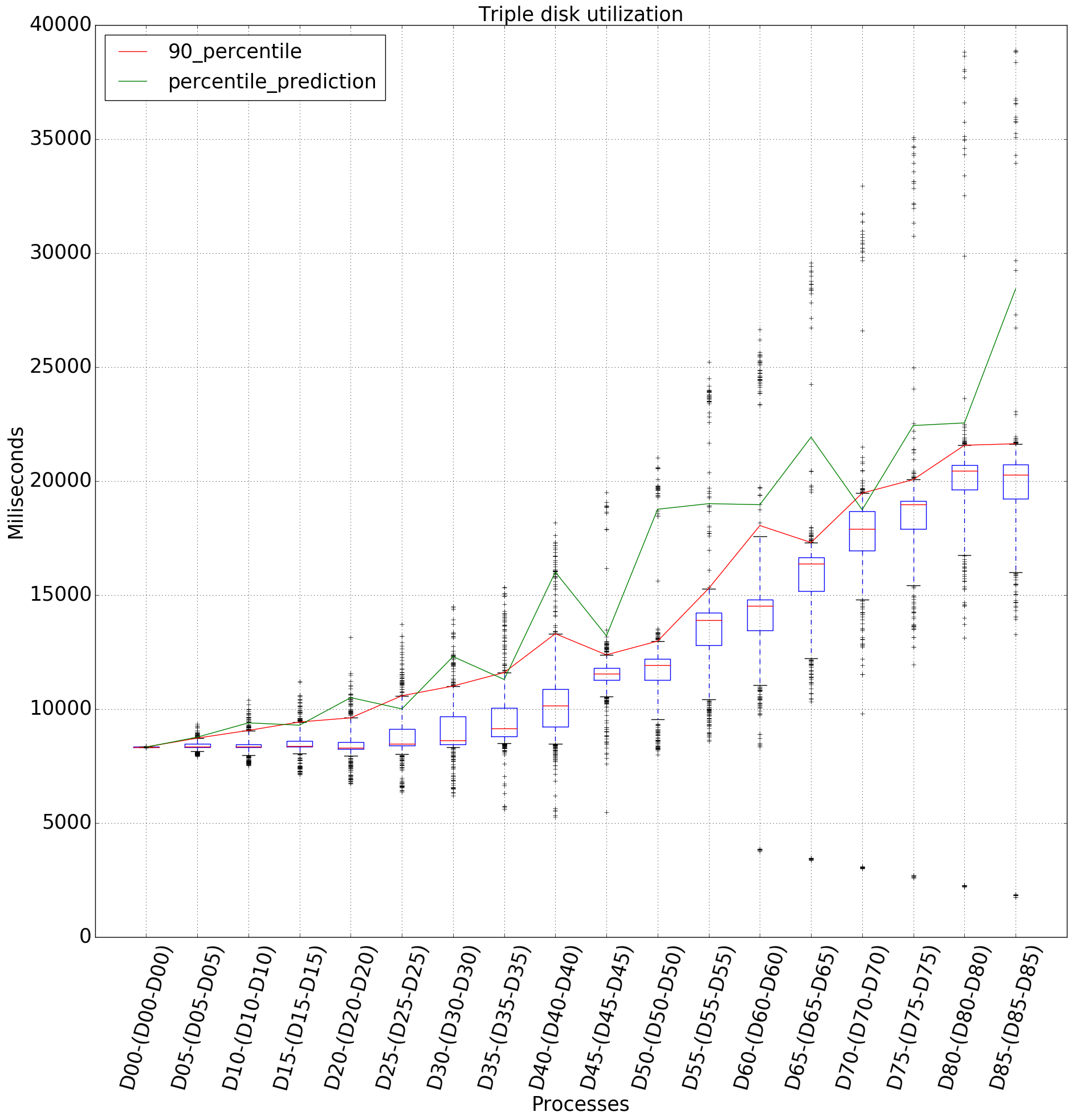}
    \caption{Benchmark response time for 3 colocated microservices with varying levels of IO utilization.}
    \label{fig:disk_paper}
\end{figure}

The results of the experiment are shown in Figure~\ref{fig:disk_paper}.
The points on the X axis correspond to IO utilization in each of the 3 containers, while the Y axis shows the response time.
We can observe that until 20\% container IO utilization (60\% overall), the median and both quartiles of the response time remain stable, while the extremes grow linearly with increasing utilization.
At 25\% container IO utilization (75\% overall), the response time inter-quartile range becomes wider as the microservices start to influence each other and the variance of the response times increases.
At 35\% container IO utilization (105\% overall), the median response time starts to increase, and grows (along with the extremes) in a linear fashion as the overall demand for IO bandwidth exceeds the available IO capacity.

The predictor provides relatively conservative estimates of the response time 90th percentile (above or slightly below the observed 90th percentile) throughout the entire range.
This suggests that the predictor's operational range could include situations in which the overall IO demand exceeds the available IO capacity, possibly depending on the workload.
We therefore conservatively limit the predictor's operational boundary to 100\% of the available IO capacity.

\subsubsection{Memory utilization}
In this final experiment we focus on the interactions between memory-bound microservices, and analyze how their response times change as their memory utilization increases.
We define memory utilization as a fraction of time a microservice spends on memory-bound tasks---allocating, reading, and writing into memory.
Note that while we do not explicitly define the available (system-wide) memory utilization capacity, we assume that a single service spending all of its processing time performing the memory-intensive operation can (almost) saturate the memory subsystem.
Similarly to the previous experiments, we deploy 3 microservices simulated by our benchmark tool to 3 containers, and vary their memory utilization from 0\% to 100\%, resulting in overall memory utilization demand ranging from 0\% to 300\%.

\begin{figure}
    \includegraphics[width=\linewidth]{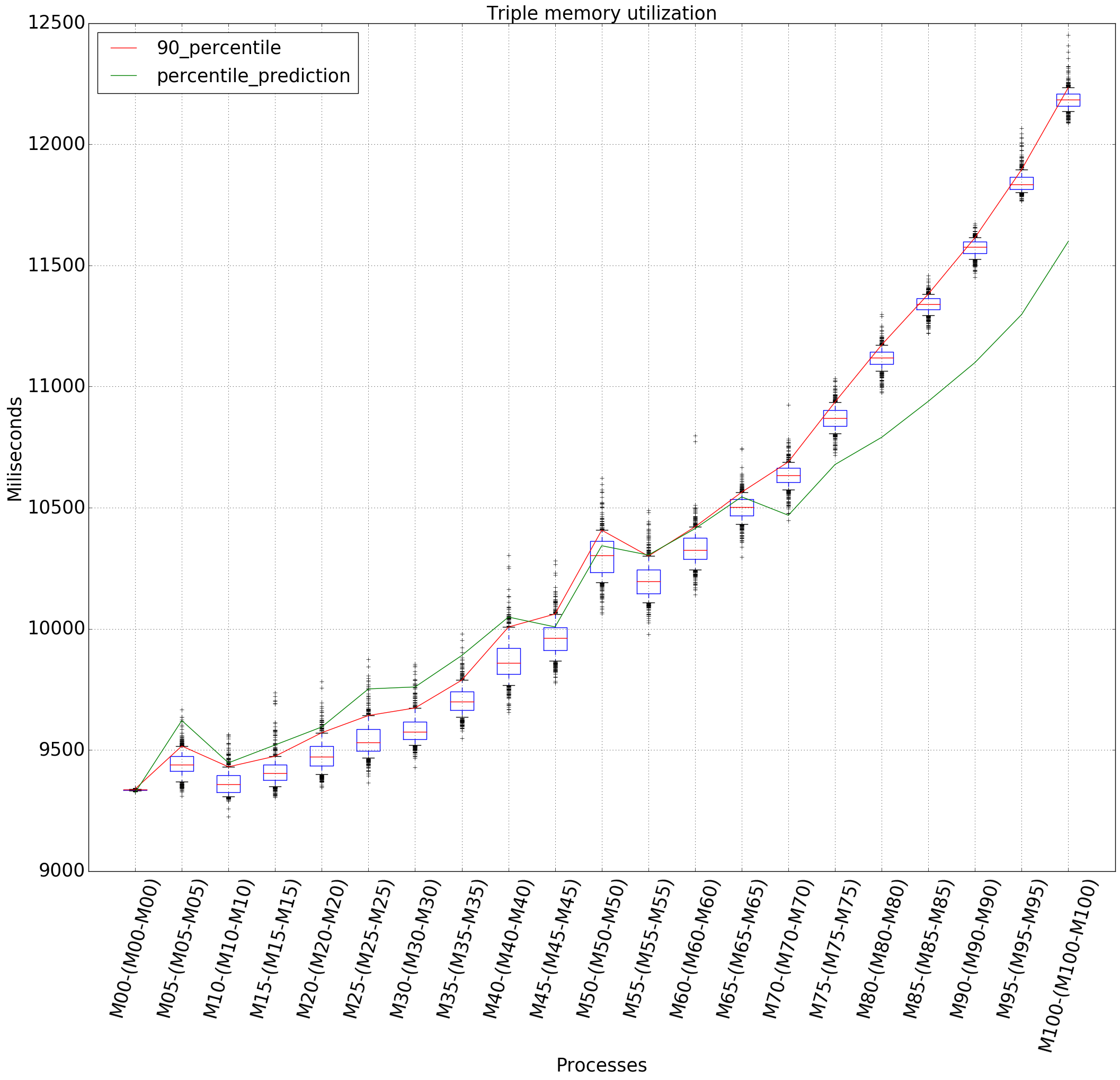}
    \caption{Three concurrent processes testing memory utilization}
    \label{fig:memory_paper}
\end{figure}

The results of the experiment are shown in Figure~\ref{fig:memory_paper}.
The points on the X axis correspond to memory utilization in each of the 3 containers, while the Y axis shows the response time.
We can observe that as the memory utilization in the containers increases, the response time median and the 90th percentile increase in an exponential fashion while the inter-quartile range remains stable

The predictor provides useful estimates of the response time 90th percentile until 65\% container memory utilization (195\% overall), which is where we set the operational boundary limit of the predictor.
However, this limit is based on a synthetic benchmark in which we can control the amount of memory-intensive work performed.
This is not possible for normal workloads, therefore we need an externally observable metric (similar to IO throughput or overall CPU utilization) which we can use to ensure that the system stays within the operational boundaries.
We use the rate of last-level cache (LLC) misses per millisecond, because it was strongly correlated (growing linearly) with the memory utilization.
On the experimental platform, the 195\% overall memory utilization representing the operational boundary corresponds to approximately to an LLC miss rate of 210000/ms.

\section{Evaluation}
\label{sec:evaluation}
In this section, we evaluate the ability of the predictor to predict performance of a workload while executing together with other workloads.
Because there are no established edge-cloud application benchmarks that could be used to evaluate our approach, we opt for the next best option, which is to emulate an edge-cloud application workload using a combination of custom benchmarks and benchmarks from existing benchmark suites.

We use workloads from the Scalabench~\cite{sewe_scalabench_2011} suite (which extends the DaCapo~\cite{blackburn_dacapo_2006} suite with Scala-based workloads), and the \textsf{stress-ng}\footnote{\url{http://kernel.ubuntu.com/~cking/stress-ng/}} suite.
All the suites are well established, with the Scalabench suite providing memory- and computationally-intensive workloads executing on the Java Virtual Machine, and the \textsf{stress-ng} suite providing a variety of smaller benchmarks targeting different computational resources.
The granularity of the benchmarks from these suites also allows us to easily wrap them as edge-cloud applications that perform considerable amount of work in response to a lightweight request.
We have also implemented a number of custom benchmarks representing workloads that perform operations on commonly used data structures, JSON processing, compression, or face recognition.

This gives us a relatively wide gamut of 17 benchmarks which reflect workloads that are likely to be found in edge-cloud applications.
The benchmarks are listed in Table~\ref{tab:selected-benchmarks}, categorized based on their resource usage.

\begin{table*}[!b]
  \renewcommand{\arraystretch}{1.2}%
  \begin{tabularx}{\textwidth}{|l|l|X|l|l|l|}
    \hline
    \multirow{2.3}{*}{\textbf{Benchmark ID}} & \multirow{2.3}{*}{\textbf{Source group}} & \multirow{2.3}{*}{\textbf{Benchmark description}} & \multicolumn{3}{c|}{\textbf{Resource Demand}} \\ 
    \cline{4-6}                              &                                          &                                                   & \textbf{CPU} & \textbf{RAM} & \textbf{Disk}   \\

    \hline
    A       & scalabench & Renders a set of images using ray tracing                         & ++++ & +    & *    \\ 
    \hline
    F       & scalabench & In-memory benchmark of transactions in banking application        & ++   & ++   & *    \\ 
    \hline
    H       & scalabench & Framework to optimize ABC, SWC, and SWF files                     & ++   & +++  & *    \\ 
    \hline
    K       & scalabench & Stanford Topic Modeling Toolbox                                   & +++  & ++++ & *    \\ 
    \hline
    O       & scalabench & Simulates programs run on a grid of AVR microcontrollers          & ++   & +    & *    \\ 
    \hline
    SMATRIX & stress-ng  & Transposition on a 4096x4096 matrix                               & +    & ++++ & *    \\ 
    \hline
    JSOND   & own        & Generates and writes JSON data to disk                            & ++   & +    & +    \\ 
    \hline
    PDFD    & own        & Generates images and writes them as PDF file to disk              & +    & ++   & +++  \\ 
    \hline
    SORTD   & own        & Generates, sorts and writes random numbers to disk                & +    & ++   & ++   \\ 
    \hline
    CYPHERD & own        & Generates random string, cyphers it and writes to disk            & +    & ++   & ++   \\ 
    \hline
    AVL     & own        & Inserts and then removes 1 000 000 items to AVL tree              & +    & ++   & *    \\ 
    \hline
    RB      & own        & Inserts and then removes 1 000 000 items to Red--Black tree       & +    & ++   & *    \\ 
    \hline
    FLOYD   & own        & Floyd-Warshall's all pairs shortest path search on 2 200 vertices & +    & +++  & *    \\ 
    \hline
    ROD     & own        & Rod cutting problem using dynamic programming                     & ++   & ++++ & *    \\ 
    \hline
    EGG     & own        & Egg dropping problem using dynamic programming                    & +    & +++  & *    \\ 
    \hline
    FACE    & own        & Human face detection in images from the local directory           & +    & +++  & +    \\ 
    \hline
    ZB      & own        & Zip archive extraction of compressed folder with many small files & +    & +    & ++++ \\
    \hline
  \end{tabularx}
  \caption{List of selected benchmarks. The meaning of the + symbol is different for each of the columns.
  For the CPU column, every + represents an additional 25\% of commenced total CPU usage.
  In the RAM column, the first + corresponds to at most 35~000 last level cache misses per millisecond of execution, each additional + represents additional 70~000 LLC misses/ms.
  In the Disk column, the * symbol represents negligible disk usage (<5\%), the first + symbol represents 5\%-25\% disk utilization, and each additional + represents additional 25\% of disk utilization.}
\label{tab:selected-benchmarks}
\end{table*}

Our experimental evaluation models a situation in which the scheduler uses the predictor to determine whether a new workload can be admitted to the system.
Benchmarks are used to represent both the workloads already executing in the system and the workload being subjected to admission gating.
To predict the performance of the workload being admitted, the predictor uses measurement data from isolated execution of each of the workloads (both already executing and the one being admitted), as well as data from combinations of a subset of the executing workloads with the workload being admitted.
In summary, the scheduler does not have complete information (i.e., it does not have measurements reflecting simultaneous execution of all the workloads), and uses the predictor to determine whether the system will be still able to provide adequate level of service if the new workload is admitted to the system.

Given the number of benchmarks evaluated, the number of benchmark combinations is too high and the results are too numerous to present in an article.
We therefore present results for combinations of the four most representative benchmarks and make the results of evaluation with other benchmarks available online\footnote{\url{https://smartarch.github.io/jss-2019-benchmark-results/}}.

The results presented in this section are therefore limited to the following benchmarks:
\begin{itemize}
  \item A CPU intensive application (referred to as ``A'') is represented by the \emph{Sunflow} benchmark from the DaCapo suite (version 9.12-bach).
  Sunflow utilizes a multi-threaded global-illumination rendering system for photo-realistic image synthesis.

  \item A disk intensive application (referred to as ``ZB'') is represented by the \emph{unzip tool}\footnote{\url{https://github.com/zeroturnaround/zt-zip}}.
  The application extracts a 188MB zip file containing 4396 files and 184 directories, producing a total of 1.08GB of uncompressed data written to disk.

  \item A memory intensive application (referred to as ``SM'') is represented by the \emph{matrix} benchmark from the \textsf{stress-ng} suite.
  The application repeatedly performs transposition of a 4096x4096 matrix.

  \item A CPU and memory intensive application (referred to as ``FACE'') is represented by the \emph{JJIL} library\footnote{\url{https://code.google.com/archive/p/jjil/}}.
  The application detects human face on 148 input images and generates the same amount of output images overlaid with a mask that identifies the detected faces.
\end{itemize}

In the following experiments, we evaluate the accuracy of the prediction algorithm in two scenarios corresponding to how our edge-cloud scheduler uses the algorithm.
Specifically, we compare a predicted and a measured percentile of the response time of a workload executing together with a combination of other workloads.

While it would be unreallistic to expect the prediction algorithm to be completely accurate, of the two possible kinds of inaccuracy we would prefer the predictor to make conservative estimates (predicting longer response times than measured) so that the scheduler can provide response time guarantees and ensure the system does not leave operational boundaries of the prediction algorithm.
At the same time, the prediction must not be too conservative, because that would prevent the scheduler from colocating workloads, leading to a system that meets (probabilistic) response time guarantees, but does not utilize resources efficiently.

In the first scenario, the scheduler measures performance for all benchmarks executing individually, and for all combinations of pairs of colocated benchmarks.
The scheduler then uses the measured data to predict the performance of a third benchmark executing together with a specific pair, which represents a situation in which an application (the third benchmark) is being considered for admission into the system.

\begin{figure}
    \includegraphics[width=\linewidth]{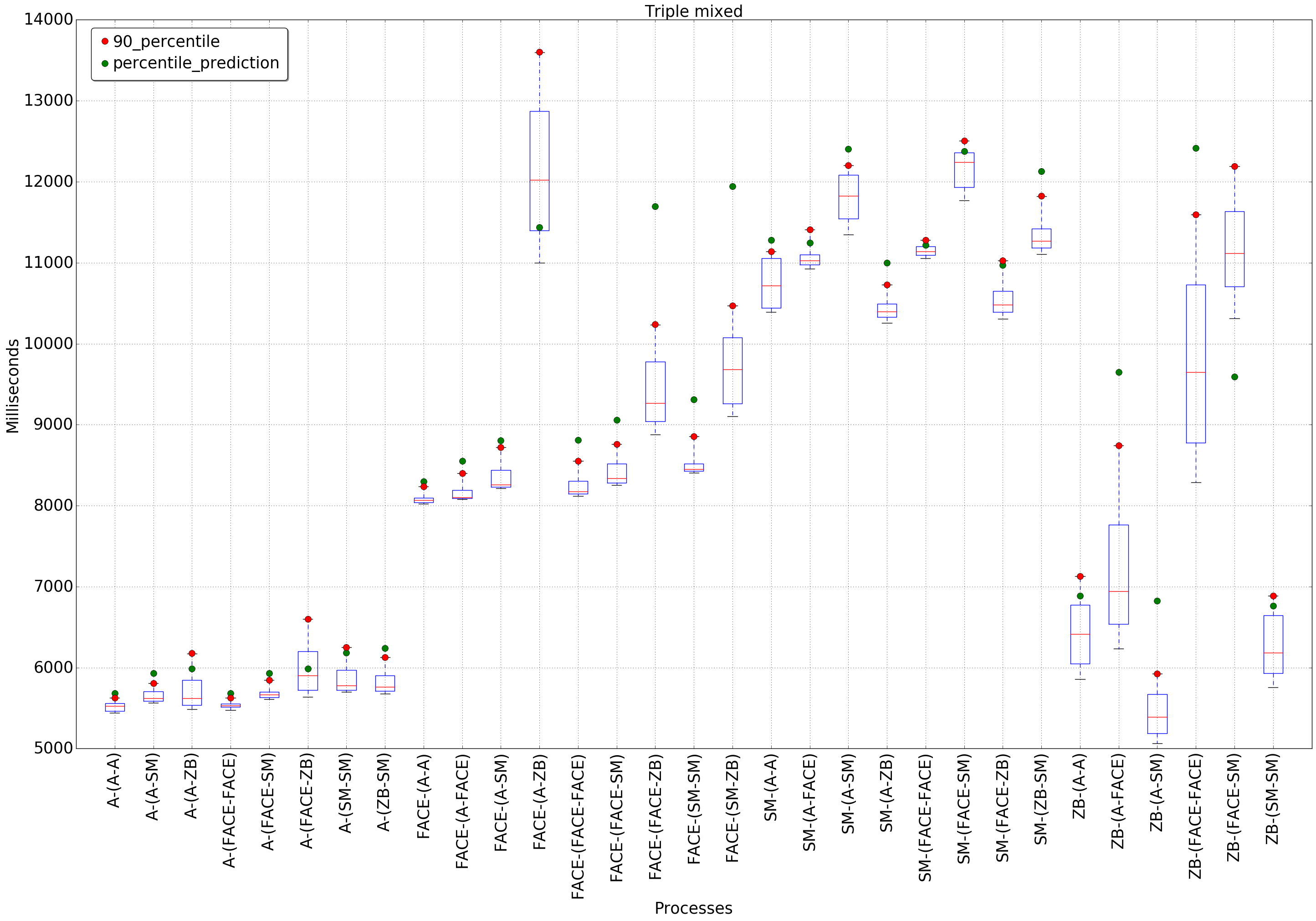}
    \caption{Evaluation of prediction algorithm predicting triplets.}
    \label{fig:triplets_sorted}
\end{figure}

The results for the first scenario are shown in Figure~\ref{fig:triplets_sorted}.
The values on the X-axis represent different combinations of workloads.
Enclosed in parentheses are the already-deployed pairs of workloads for which the scheduler has performance data available.
The values on the Y-axis correspond to response time of the third benchmark being deployed.
The red circles in the plot represent the measured 90th percentile of the benchmark response time, while the green circles represent the predicted value of the percentile.
We can observe that in 20 out of 31 configurations the predictor estimates the response time 90th percentile conservatively (higher than measured), and stays within a relative range of [-21\%, +15\%] below/above the measured value, with an average of +4.9\%.

\begin{figure}
    \includegraphics[width=\linewidth]{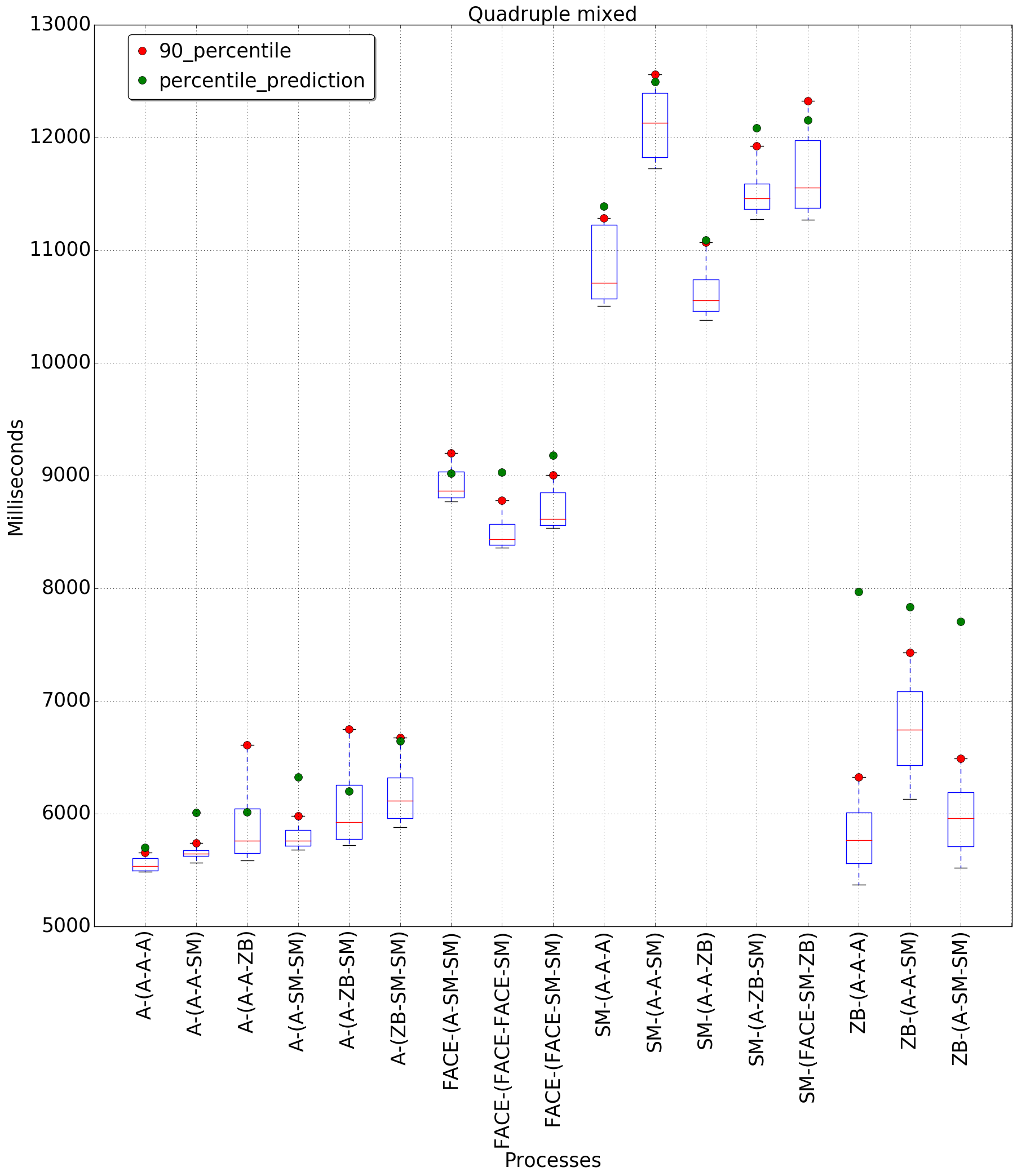}
    \caption{Evaluation of prediction algorithm predicting quadruplets.}
    \label{fig:quadruplets_sorted}
\end{figure}

In the second scenario, the scheduler also measures performance for all combinations of three colocated benchmarks, and uses the measured data (single, pairs, and triplets) to predict the performance of a fourth benchmark being colocated with a particular triplet.
The results for the second scenario are shown in Figure~\ref{fig:quadruplets_sorted}.
We can observe that in 11 out of 17 cases the predictor estimates the response time 90th percentile conservatively, and stays within a relative range of [-10\%, +26\%] below/above the measured value, with an average of +5.3\%.

The results exclude combinations containing more than one instance of the ``ZB'' benchmark, because such a configuration falls outside the operational boundary of the predictor---a single instance of the ``ZB'' benchmark utilizes 65\% of the IO bandwidth.
Similarly, we excluded combinations containing three or more instances of the ``SM'' benchmark because they fall far outside the operational boundary of the predictor expressed by the number of last-level cache misses per millisecond---the limit is 210000/ms, while three and four instances of the ``SM'' benchmark exhibit an LLC miss rate of 450000/ms and 600000/ms, respectively.

\section{Discussion}
\label{sec:discussion}

In general, the problem of managing QoS in a cloud is extremely complex, providing many problem facets, each sufficient to sustain an entire research area.
Consequently, the amount of existing work is daunting, making it nearly impossible to find a solution fitting a particular context and providing usable interfaces to developers.
We therefore take advantage of the edge-cloud context to avoid some of the complexities traditionally associated with QoS management in public clouds, where a generic solution has to deal with unknown applications, unknown workloads, unknown developers, and unknown clients.

We are aware that the operational boundaries of the prediction algorithm may depend on the particular combination of colocated workloads.
Therefore in general, the operational boundaries themselves may need to be predicted as well, which goes beyond the scope of this paper.
However, in this particular context, we assume that the set of possible microservices or microservice types for a particular (privately controlled) edge-cloud infrastructure will be somewhat restricted (with a bias towards latency sensitive applications).
We have therefore chosen approximate static boundaries determined using synthetic benchmarks and leave the prediction of the operational boundaries to future work.
A restricted set of possible microservices or microservice types also provides some ``combinatorial headroom'' in the staging phase where the resource demand and performance interference between combinations of microservices is evaluated.

The IO utilization in our model of the edge-cloud microservice demands does not include network bandwidth, and we consider it unlimited for modeling purposes.
The rationale for this decision is that edge-cloud applications are very likely to be latency-sensitive, but not bandwidth-intensive---that would defeat the primary purpose of edge-cloud which is to reduce communication latencies due to distance.
We therefore assume that the network infrastructure can be configured to assign time-critical network traffic a QoS class with high priority and that latency-sensitive services with response time requirements will not saturate the network with bulk transfers.

\section{Related work}
\label{sec:related-work}
Cloud computing has been both a blessing and a curse.
Cloud users can benefit from unprecedented availability and elasticity of resources, but the benefits come with strings attached.
Cloud infrastructure and service providers have to continually balance the tension between efficient resource utilization (which determines costs) on the one hand, and quality-of-service guarantees demanded by providers of latency-sensitive (LS) applications on the other hand.
Management of cloud resources has therefore become a vast and quickly moving research area, with many surveys mapping and categorizing the problems, challenges, and the state-of-the-art in various problem domains \cite{chen_survey_2018,amiri_survey_2017,hameed_survey_2016,singh_qos-aware_2015,faniyi_systematic_2015,mann_allocation_2015,garcia-valls_challenges_2014}.

To position our work in the context of other research in the area, we first review some of the works (ordered chronologically) that we consider most relevant to our approach and then discuss specific points that define the frame of reference for our research.
Given the amount of existing research in this area, our selection is admittedly far from exhaustive.

Q-Clouds~\cite{nathuji_q-clouds_2010} is a QoS-aware control framework which transparently adjusts resource allocation to mitigate effects of interference on shared resources.
Q-Cloud first profiles the virtual machines (VM) submitted by clients on a staging server to assess the amount of resources needed to attain the desired QoS without interference, and then manages the resources allocated to the deployed VMs in a closed control loop.

Cuanta~\cite{govindan_cuanta_2011} is a technique for predicting performance degradation due to shard processor cache for any possible placement using a linear (as opposed to exponential) number of measurements.
Applications are replaced by a synthetic clone which is tuned to mimic the application's cache pressure, and interference due to colocation is predicted based on a matrix of know interference effects between different configurations of cache clones.
Even though Cuanta is not a full-fledged cloud scheduler, it was used to make better workload placement decisions for a given performance and resource constraints.

Bubble-Up~\cite{mars_bubble-up_2011} avoids pairwise colocation profiling by characterizing the QoS degradation in LS applications using a synthetic workload with configurable memory subsystem stress test (the \emph{bubble}), and the contentiousness of batch applications using a \emph{reporter} workload with known sensitivity curve.
The contentiousness of a batch application is mapped to a configuration of the bubble, which is then used to predict the interference inflicted by the batch application on the LS application.
Bubble-Flux~\cite{yang_bubble-flux_2013} improves on Bubble-Up by performing online profiling for LS workloads to account for workload phase changes and to identify more colocation opportunities.

Paragon~\cite{delimitrou_paragon_2013} is an online interference-aware scheduler which uses collaborative filtering to classify incoming applications based on limited profiling signal and similarity to previously scheduled applications.
It does not differentiate between batch and LS applications and schedules applications so as to minimize interference and maximize utilization.
Applications are classified for interference tolerance using microbenchmarks stressing a specific shared resource with tunable intensity, which are run concurrently with an application to find out the interference level at which the application's performance falls below 95\% of its performance in isolation.
Quasar~\cite{delimitrou_quasar_2014} improves on Paragon in that it also performs resource allocation instead of only resource assignment.
Quasar extends the classification engine of Paragon to consider scale-out and scale-up scenarios, as well as different workload types with different constraints and resource allocation controls.
It also provides an API that allows expressing the performance constraints regarding throughput and latency.

CloudScope~\cite{chen_cloudscope_2015} is a representative of model-based approaches to QoS-aware cloud resource management and uses a discrete-time Markov Chain model to predict performance interference of colocated VMs.
CloudScope runs within each host and collects application and VM-related metrics at runtime.
The metrics serve to maintain an application-specific model capturing the proportion of the time an application uses a particular resource.
The model is then used to predict slowdown due to colocation and ultimately to control placement of guest VM instances as well as adjusting the resources available to a hypervisor.

CtrlCloud~\cite{adam_ctrlcloud_2017} is a performance-aware cloud resource manager and controller, which optimizes the allocation of CPU resources VMs to meet QoS targets.
It maintains an online model of the relationship between allocated resource shares and the application performance, and uses a control loop to adapt the resource allocation so as to progress towards a probabilistic performance target expressed as a percentile of requests that must observe a response time within certain bounds.

Pythia~\cite{xu_pythia_2018} is a colocation manager which uses a linear regression model to predict combined contention on shared resources when colocating multiple batch workloads with an LS workload.
Pythia performs contention characterization for each batch workload running together with a particular LS workload and removes batch workloads that are too contentious to allow safe colocation.
It then selects a small subset of batch workloads to colocate with a latency sensitive workload and measures their combined contention to build a linear regression prediction model for contention due to multiple batch workloads.

\smallskip
In general, the approaches presented above, including our own, aim to provide a performance- and interference-aware self-adaptive system (an essential part thereof) which manages resource allocation and assignment in a cloud environment to achieve efficient utilization of available resources while allowing applications to meet their QoS target.
Our selection illustrates the variety of approaches proposed over the years, each fitting a different context, but none able to claim to solve the problem once and for all---our approach is not different.

Similarly to Pythia, our approach profiles applications in a staging area, but does not utilize proxy workloads like most of the other approaches.
We profile the application-provided probes to determine resource demands for representative workload, but also periodically monitor the response time on these probes to ensure that the current deployment provides the desired performance guarantees, which is again similar to other approaches, with the exception of Bubble-Up and Cuanta.
This is important because by using dedicated probes, we do not force developers to mix processing of proxy workloads with regular user requests in the business logic of an application.
Also, the use of dedicated probes, over which we provide guarantees, gives more precise semantics to the guarantees and creates a suitable contract between the application developer and the cloud, which treats the applications as black boxes.

Unlike other approaches, our approach to dealing with performance interference treats all resources equally and relies on statistical characterization and similarity to reveal dependencies between background workloads.
We are aware of potential non-linear relationships which might be difficult to predict using our model, and therefore actively limit and most importantly enforce the operational boundaries of our prediction algorithm.
This is where the inherent integration of the prediction method with the self-adaptation mechanism creates novelty in our approach, as we not only use the prediction to control the admission, but we also control admission to preserve the quality of the prediction.

We specifically target (non-public) edge-cloud environment, which allows us to constrain the problem and cater to context-specific details.
Instead of VMs, we focus on container technologies and on providing probabilistic performance guarantees.
We require the developers to provide monitoring probes and to explicitly specify performance objectives for these probes, and only admit applications for deployment if the system considers the objectives satisfiable.
Other than that, we treat each application as a black box.
Our focus on containers stems mainly from their lower overhead and higher flexibility (compared to VMs), which allows us to relocate services more flexibly in response to mobility of end-users.

\smallskip
Another related area is represented by works targeting \textit{service-level agreements} (SLAs) in cloud environment, including edge-cloud.
In general the main difference between our approach and classic SLAs for clouds is that we target a different type of edge-cloud (non-public, privately-controlled), which allows us to make additional assumptions about the underlying infrastructure.
Another difference is that our approach is based on estimations of worst-case response time (to provide soft-realtime guaranties), while other approaches focus on estimating throughput.
Finally, our SLAs are interconnected with adaptation which is not retroactive but proactive, i.e., we build a performance model of a microservice from measurements and can therefore react (proactively) to different situations.

In the following, we review some of the approaches that differ from classic SLAs and bear more similarity to our approach.

The work of Remesh et al. \cite{remesh_babu_servicelevel_2019} deals with SLA-aware scheduling and load-balancing.
While the general idea is similar to ours, their primary goal is to load-balance services (i.e., re-deploy them to another computer in the cloud) in order to keep computers in the cloud evenly loaded and reduce the chance of SLA breaches.
The approach does not measure service response time, and instead relies on low-level information which has to be provided by the application developer, such as the required amount of memory, CPU speed in terms of instructions-per-second, etc.
The semantic gap between the concepts used in the SLA specification and the domain-specific application requirements makes formulation of an SLA difficult.

Cerebro~\cite{cerebro_2015} is similar to our approach in that it first statically analyzes services to find out important calls, measures the performance of these calls at runtime, and predicts bounds on response time using time-series analysis.
The main difference is that Cerebro focuses only on a specific set of calls while our approach measures the performance of a service function as represented by a probe.
Our approach does not require any other information about a service and does not rely on a fixed knowledge of what is important.
In addition, our approach uses the prediction results to manage deployment and re-deployment of services in the cloud.

Panda et al. \cite{panda_sla-based_2017} presents an SLA-based scheduling algorithm for a cloud.
Similarly to our approach, the authors consider clouds composed of multiple data centers and schedule service for deployment to data centers so as to honor the SLAs.
Contrary to our approach, the authors expect SLAs to be provided together with the services and consider only execution time, cost, and penalty for SLA violation.
The scheduling algorithm then attempts to minimize service execution time and cost.

The resource aware scheduler SLA-RALBA~\cite{RALBA} is similar to the previous approach but considers also heterogeneous clouds.

Also similar is the approach of He et al. \cite{yangui_re-deploying_2019}, in which the authors consider re-deployment of services in edge-cloud and cloud environment, but focus mainly on minimization of the re-deployment cost.

\smallskip
Finally, monitoring in the edge-cloud environment is also partially related to our work, but orthogonal in purpose.
Abderrahim et al. ~\cite{abderrahim_holistic_2017} advocate the need for a holistic monitoring services in the context of edge and fog infrastructures.
The authors analyze the requirements for such services and provide a classification and qualitative analysis of the major existing solutions.

The FMonE framework~\cite{brandon_fmone_2018} is a particular example of a dedicated edge-cloud monitoring service, which provides a flexible monitoring infrastructure for the fog computing paradigm.
Souza et al.~\cite{souza_osmotic_2018} present an edge-cloud monitoring approach (which is an extension of the CLAMBS~\cite{alhamazani_cross-layer_2019} cloud monitoring service) and which, based on monitoring, can trigger migration of micro-services between edge-nodes to improve latency or performance, but does not provide any guarantees.

In general, monitoring operates on a post-fact basis, with the primary purpose of avoiding problems going on undetected.
Monitoring solutions generally focus on indicators related to the health of systems or applications, and trigger typically coarse-grained corrective actions (restarting applications, starting or migrating virtual machines, etc).
These actions can often remedy transient problems, or stop cascading problems from occurring, but monitoring alone does not provide any execution guarantees.

While our approach shares some technical aspects with monitoring solutions, and could even make use of a monitoring solution to gather system- and application-level health information, it primarily operates at a different conceptual level.
The goal of our approach is to provide probabilistic execution guarantees, and uses monitoring data to predict application performance at probe points in different deployment scenarios.
Enacting a particular deployment scenario may trigger deployment changes in multiple managed applications, which is not a simple corrective action.

\section{Conclusion}
\label{sec:conclusion}
To summarize, we present an approach to providing soft real-time guarantees on response time of microservices deployed in an edge-cloud.
Our approach allows developers to express the desired guarantees directly in the form of probabilistic requirements (e.g., in 90\% of cases the response time should be within 100 ms).
This contrasts with solutions requiring explicit reservations in terms common for existing cloud platforms, such as CPUs or IOPS, which are disconnected from the developer's perspective of application performance.

Our approach relies on a statistical method used to predict the upper bound of a microservice response time (at a given confidence level) when colocated with other microservices.
Based on this prediction, an adaptive control loop can manage the deployment and re-deployment of microservices to proactively maintain the performance targets.

Because the effects of performance interference between colocated workloads can be non-linear (especially when resources are overcommitted) and can negatively impact the accuracy of the response time predictions, we limit the operational boundary of the prediction algorithm to system states in which the algorithm provides good estimates, which can be broadly described as non-overcommitted.
Consequently, the edge-cloud scheduler controls the admission and deployment of microservices not only to provide response time guarantees, but also to ensure that the managed systems stay within the operational boundary of the prediction algorithm.
It is important to mention that non-overcommitment does not present any issue as we primarily consider microservices with continuous workloads (stream processing, etc.~--~as already mentioned in the introduction).
While overcommitment makes sense in the case of microservices with processing bursts (and longer periods of inactivity between the bursts), it provides no benefits in our case.
Because the workload is continuous, overcommitting automatically implies the inability to meet the timing requirements.

An important novelty of our approach (with respect to the state of the art) is that it has minimal impact on microservice developers.
In our approach, we treat the microservices as black boxes that the framework knows nothing about---apart from performance requirements expressed over application probes.
The edge-cloud infrastructure automatically profiles the microservices submitted for deployment and collects the data needed for the prediction algorithm.
Consequently, we do not impose any specific software architecture (such as partitioning into real-time tasks) on the developer, nor do we require any particular programming language to be used.
More importantly though, we do not require the developer to provide a performance model of the application, or specify the application resource requiremnts in terms of CPUs and IOPS, which are disconnected from his or her perception of application performance.

Even though in this paper we do not specifically focus on networking, in future we plan to include the knowledge of network topology and network latencies in the control loop to enable more flexible distribution of microservices across edge-cloud nodes.

\section*{Acknowledgment}

The research leading to these results has received funding from the ECSEL Joint Undertaking (JU) under grant agreement No 783162, where the method has been adapted and applied for processing video workloads. 
Also, it has received funding from the ECSEL Joint Undertaking (JU) under grant agreement No 783221, where the method has been adapted and applied for processing farming related workloads. 
Additionally, this work was partially supported by Charles University institutional funding SVV 260451.

\bibliographystyle{model1-num-names}

\bibliography{paper-jss-journal,related}

\end{document}